\documentclass[a4paper,11pt]{article}
\usepackage{pos}
\usepackage[english]{babel}		
\usepackage{graphicx}			
\usepackage{hyperref}
\usepackage{subcaption}			

\title{Quality assurance of actuators for the Medium-Sized Telescopes of the Cherenkov Telescope Array}
\ShortTitle{Quality assurance of actuators for the Medium-Sized Telescopes of the CTA}

\manuallySeparateAuthors
\author*[a]{H.~Salzmann}
\author[a]{, J.~Dick}
\author[a]{, S.~Diebold}
\author[a]{, G.~P\"uhlhofer}
\author[a]{, S.~Renner}
\author[a]{ and A.~Santangelo}
\author{ for the CTA MST Project}

\affiliation[a]{Institut f\"ur Astronomie und Astrophysik, Eberhard Karls Universit\"at T\"ubingen\\
  Sand 1, 72076 T\"ubingen, Germany}

\emailAdd{heiko.salzmann@astro.uni-tuebingen.de}

\abstract{The Cherenkov Telescope Array (CTA) is a future ground-based observatory for gamma-ray astronomy providing unparalleled sensitivity in the energy range from $20\,$GeV up to $300\,$TeV. CTA will consist of telescopes with three different sizes. The Medium-Sized Telescopes (MSTs) will have $12\,$m reflectors with a tessellated mirror design of $86$ mirror facets each. Each mirror facet is mounted on the mirror support structure with two actuators that are adjustable in length to align the mirrors, and a freely rotating fixpoint. Image resolution and pointing accuracy constraints impose limits on the backlash and deformation of the actuators and the fixpoint under various weight and wind loads. In this contribution, the test stand to measure the backlash and deformation behaviour of actuators and fixpoints is described and the measurement procedure is explained.}

\FullConference{%
  7th Heidelberg International Symposium on High-Energy Gamma-Ray Astronomy (Gamma2022)\\
  4-8 July 2022\\
  Barcelona, Spain\\}

\begin{document}
\maketitle

\section{The Mirror Design of the Medium-Sized Telescope}
\begin{figure}[htbp]
	\centering
	\includegraphics[width=0.7\textwidth]{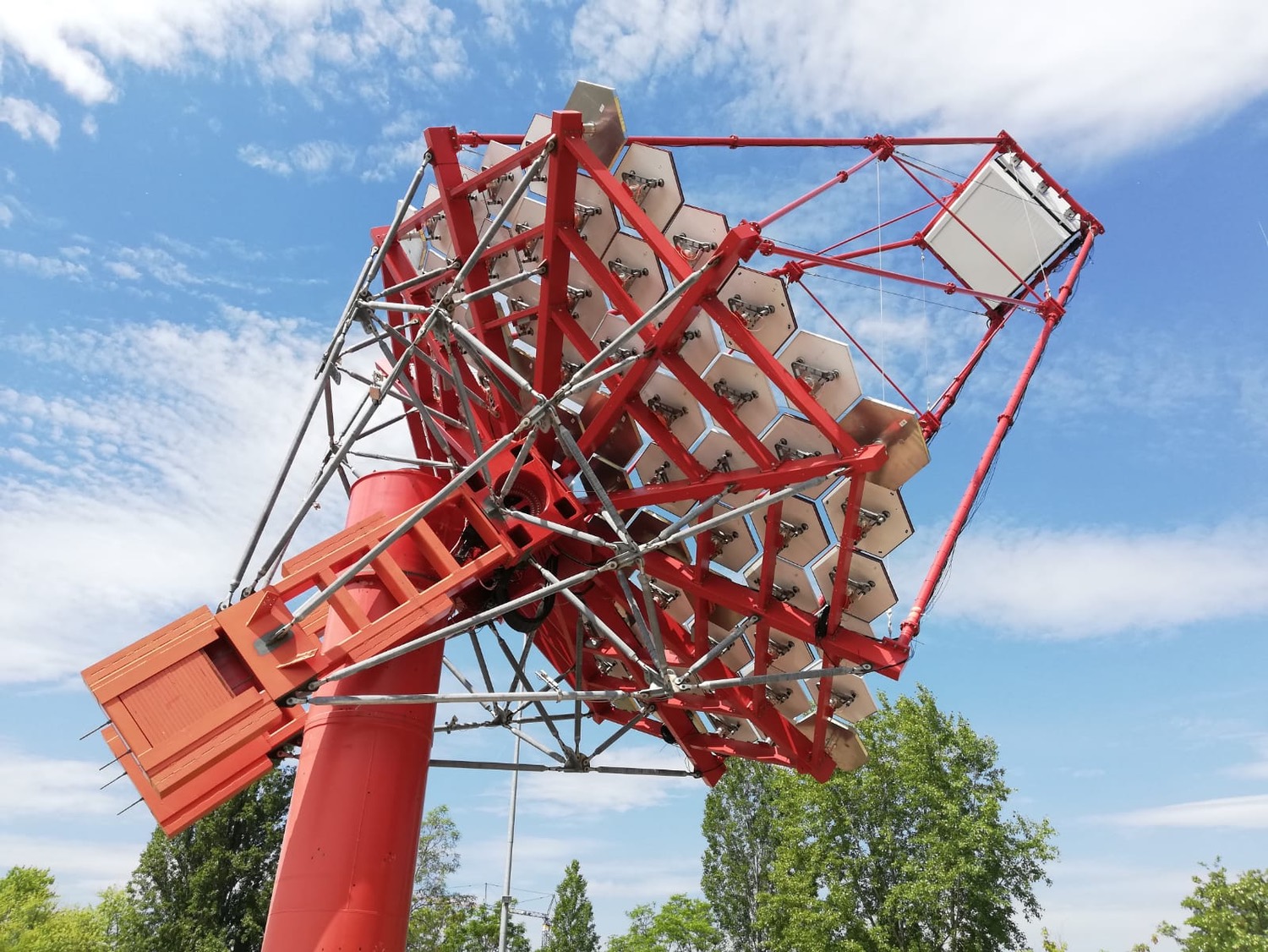}
	\caption{The MST prototype in Berlin Adlershof. Image credit: CTAO}
	\label{sec01figMST}
\end{figure}

The Cherenkov Telescope Array (CTA) is a future ground-based observatory for gamma-ray astronomy covering an energy range from $20\,$GeV up to $300\,$TeV. Three different-sized telescope types are planned for CTA, which will be located at two sites: La Palma in the northern hemisphere (CTA-North) and Chile in the southern hemisphere (CTA-South). The MST covers the energy range from $150\,$GeV to $5\,$TeV and is therefore essential for the Key Science Projects (KSPs) of CTA \cite{ctascience}. Furthermore, it is the only telescope type that will be deployed at both CTA sites for the Alpha configuration, which comprises 9 MSTs at CTA-North and 14 MSTs at CTA-South. An MST prototype is displayed in \autoref{sec01figMST}.\\
The MST is a $12\,$m modified Davies-Cotton reflector with a tessellated mirror design consisting of $86$ mirror facets. The mirror facets have a flat-to-flat diameter of $1.2\,$m and each is mounted on the mirror support structure with two motorized actuators and a fixpoint, which constitute an actuator set \cite{puehl2017}. The fixpoint is freely tilting and the actuators have an adjustable stroke length to tilt the mirror facets. This allows the mirror facets to be aligned such that the incoming light of a point source is reflected by each mirror facet into the same position in the focal plane. The precise adjustment of the mirror facet tilts is paramount for the pointing and image resolution qualities of the telescope.  

\section{Quality Assurance of Actuator Sets}
\begin{figure}[htbp]
	\centering 	
	\includegraphics[width=0.75\textwidth]{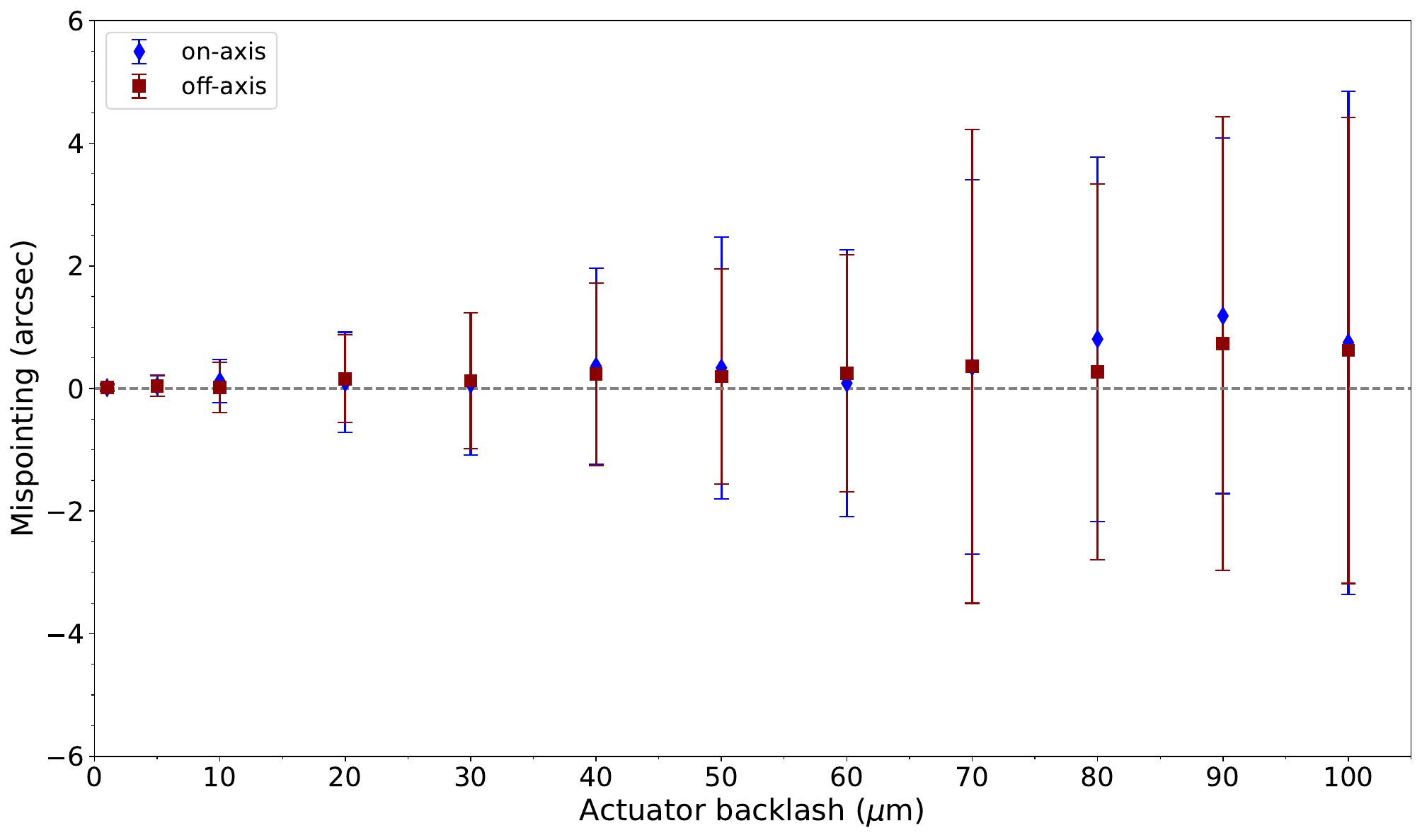}
	\caption{Impact on the on-axis (blue) and off-axis (red) pointing accuracy as determined with ray-tracing simulations assuming a randomized actuator backlash between $\pm 100\,\mu$m for a fully equipped MST. Hereby, on-axis is the light that falls into the telescope's reflector parallel to the telescope's optical axis, whereas off-axis light has an incident angle - in this case of $2.8\,\mathrm{deg}$ - to the telescope's optical axis. Image credit: Adapted from M. Garczarczyk (DESY Zeuthen)}
	\label{sec02raytracing}
\end{figure}

The absolute mirror facet tilt is influenced by two aspects.\footnote{Only deformations in the actuator stroke direction are considered and the elastic bending of the actuators and fixpoint perpendicular to the actuator stroke direction is neglected here as its influence on the mirror tilt (specifically relative to a mean bending) is small, because it primarily only leads to a plane-parallel shift of the individual mirror facets.} Firstly, its accuracy is limited by the finite resolution of the actuator stroke position on the order of $\pm5\,\mu$m defined by the actuator step size. Secondly, the reproducibility of the tilt is influenced by the backlash, the elastic and plastic deformation of the actuators as well as of the fixpoints. Furthermore, the reproducibility is depending on the observation conditions like different wind loads on the mirrors and varying weight loads due to changing telescope elevation angles. Consequently, the requirements on the image resolution and the telescope pointing accuracy impose limits on the backlash, elastic and plastic deformation of the actuators and fixpoints. Therefore, the quality assurance of actuators and fixpoints plays an important role to verify the limits on the backlash and the deformation of the actuators and fixpoints in order to fulfil the requirements on the MST, which are:
\begin{itemize}
	\item image resolution: $80$\% light containment diameter $\Theta_{80\%} < 0.18\,\mathrm{deg}$ up to $2.8\,\mathrm{deg}$ from the centre of the camera field of view\footnote{The containment diameter $\Theta_{80\%}$ is the angular diameter within which $80$\% of the light reflected by all mirror facets into a $1\,\mathrm{deg}$ diameter circle in the camera falls. Furthermore, this requirement constrains the $\Theta_{80\%}$ also for off-axis light, i.e. light that has an incident angle towards the telescope's optical axis, up to $2.8\,\mathrm{deg}$.}, and 
	\item pointing accuracy: $<7''$.
\end{itemize}
The impact of a randomized actuator backlash within $\pm 100\,\mu$m on the pointing accuracy for a fully equipped MST as derived from ray-tracing simulations is shown in \autoref{sec02raytracing}. It should be noted that the requirement on the pointing accuracy dominates the limits on the backlash and the deformation of the actuators and fixpoints.  
\\
The Institut f\"ur Astronomie und Astrophysik T\"ubingen (IAAT) possesses expertise from the development and testing of a proposed actuator design for the CTA MST (see e.g. \cite{dick2015}). Based on this expertise, the IAAT developed a test stand for the verification of the resulting limits on the actuators and fixpoints. The objective of the test stand is to measure and quantify the backlash as well as the elastic and plastic deformation behaviour of actuators, fixpoints and single actuator components in stroke direction.

\section{Measurement Theory and Principle}
\begin{figure}[htbp]
	\centering
	\begin{subfigure}{0.6\textwidth}
		\includegraphics[width=\textwidth]{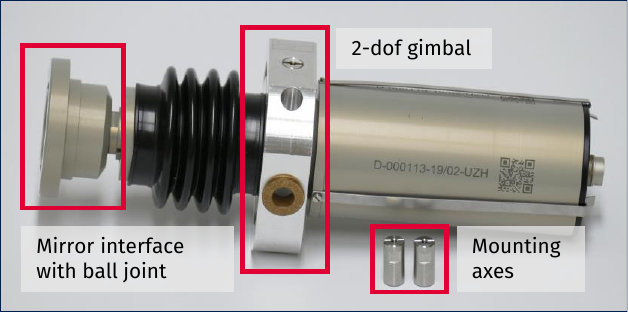}
		\label{sec02actbacklashsources}
	\end{subfigure}
	\hfill
	\begin{subfigure}{0.3\textwidth}
		\includegraphics[width=\textwidth]{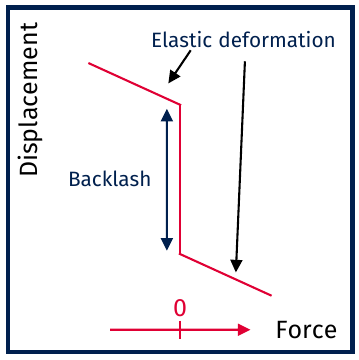}
		\label{sec02measschematic}
	\end{subfigure}
	\caption{\textbf{Left}: An actuator with a 2-degree of freedom gimbal. \textbf{Right}: Schematic sketch of the expected force-displacement diagram in the case of only backlash and elastic deformation.}
	\label{sec02combiplot}
\end{figure}

The three relevant measurement parameters to characterise the deformation behaviour of actuators and fixpoints are the backlash, elastic and plastic deformation. The backlash is a force-independent clearance between mechanical parts. The elastic and plastic deformations are force-dependent reversible and non-reversible deformations (respectively) caused by the application of an external force. In our case, non-reversible means that the deformation does not relax after the external force is removed but it only vanishes after a restoration force is applied in the opposite direction of the initial force. This leads to a hysteresis. To quantify the backlash as well as the elastic and plastic deformation for a test object, the force-displacement diagram is measured using discrete force steps. For this the maximum applied external forces are representing the various weight and wind loads present during the telescope operations. The goal is to quantify the positioning inaccuracies of the test object from the force-displacement diagram as is indicated in the schematic sketch of the force-displacement diagram for a test object only exhibiting backlash and elastic deformation on the right-hand side of \autoref{sec02combiplot}. \\
For an actuator, the possible sources of backlash are 
\begin{itemize}
	\item the actuator-internal components (e.g.\ the spindle-nut interface),
	\item the ball joint in the mirror interface, and 
	\item the gimbal. 
\end{itemize}
Furthermore, there is the possibility of backlash due to the clearance in the interface between the gimbal and the mirror support structure which is connected via the mounting axes. An exemplary actuator with a 2-degree of freedom (dof) gimbal is displayed on the left-hand side of \autoref{sec02combiplot}.

\begin{figure}[htbp]
	\centering
	\includegraphics[width=\textwidth]{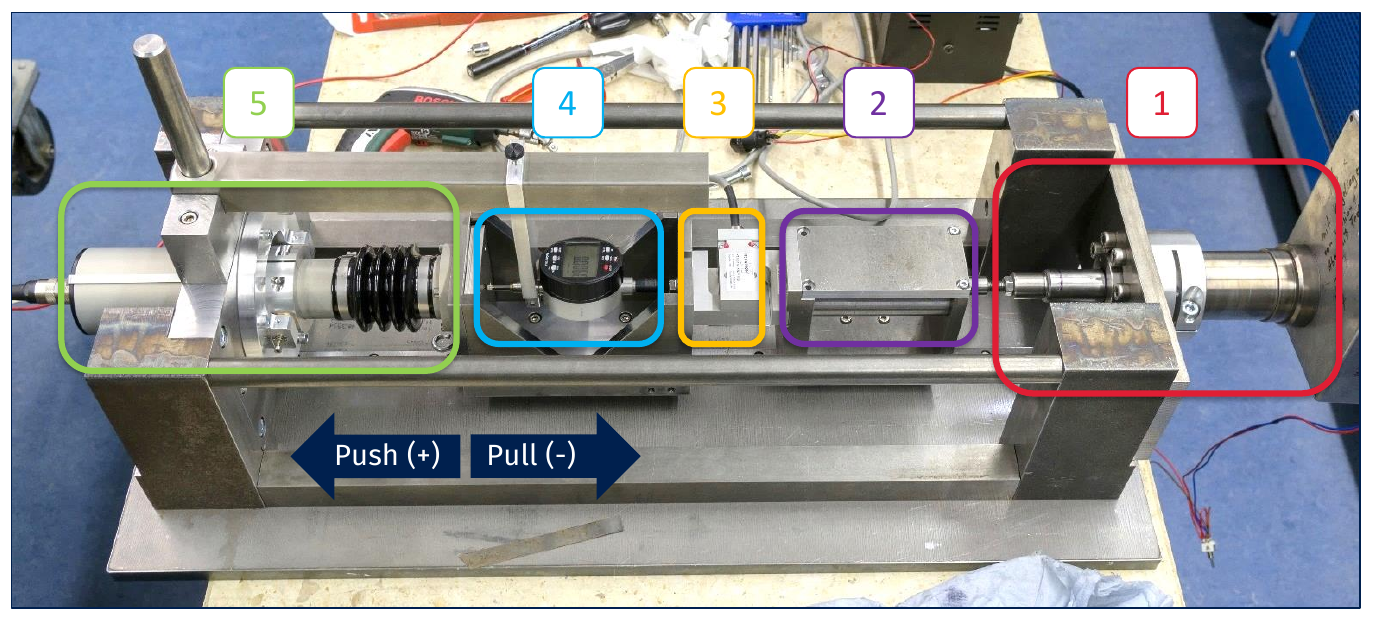}
	\caption{The actuator test stand for the actuator quality assurance. The various components are the following: 1.\ ACU, 2.\ sled with springs for linear force translation, 3.\ force sensor, 4.\ dial gauge and 5.\ test object.}
	\label{sec04teststand}
\end{figure}
\section{Test Stand for the Quality Assurance of Actuator Sets}
The test stand for the actuator quality assurance consists of five major components as is shown in \autoref{sec04teststand}. The force generation is realized with the Actuator Control Unit (ACU), which is an actuator providing the push and pull forces to compress and tense the test object, respectively. To ensure a linear relation between the ACU stroke distance and the force on the test object, the force generated by the ACU is transferred via the springs in the spring sled over the force sensor onto the sled, which is mounted to the test object. In the latter sled, the dial gauge is placed with a mounting to measure the displacement of the test object for a given force. The mounting itself is attached to the test stand. The dial gauge has a systematic measurement uncertainty of $\pm 1\,\mu$m. The test stand offers the possibility to either measure 1-dof or 2-dof actuators, as well as fixpoints and single actuator components like gimbals and ball joints. The gimbals and ball joints can be measured with a dummy system using a stiff steel rod instead of an actuator. \\
The major challenge in designing a test stand to measure the positioning inaccuracies in stroke direction of an intrinsically tilting object lies in the very precise alignment of the test stand, i.e.\ very precise alignment of the force and measurement direction. This is accomplished by connecting the ACU, the force sensor and the test object via two sleds, which are installed on a pre-tensioned guiding rail with ball bearings. The measurement of the force-displacement diagram is automatized with the software running on an external PC, which is connected to the microcontroller in the ACU and also reads out the dial gauge. The microcontroller is running the firmware realizing a feedback loop with the input of the force sensor such that the ACU can set a requested measurement force.\footnote{The feedback loop is implemented according to the ``Three level hysteresis switching with dead zone'' scheme, which is a general concept in control engineering (e.g. see Figure 2 in \cite{controlscheme}).} The accuracy of the requested measurement force is influenced by a) the accuracy of the force sensor calibration, and b) the hysteresis of the force feedback loop to avoid an oscillation of the ACU ($\Delta F_{\text{Hyst}}=\pm 8\,$N). 
\begin{figure}[htbp]
	\centering
	\begin{subfigure}{0.49\textwidth}
		\includegraphics[width=\textwidth]{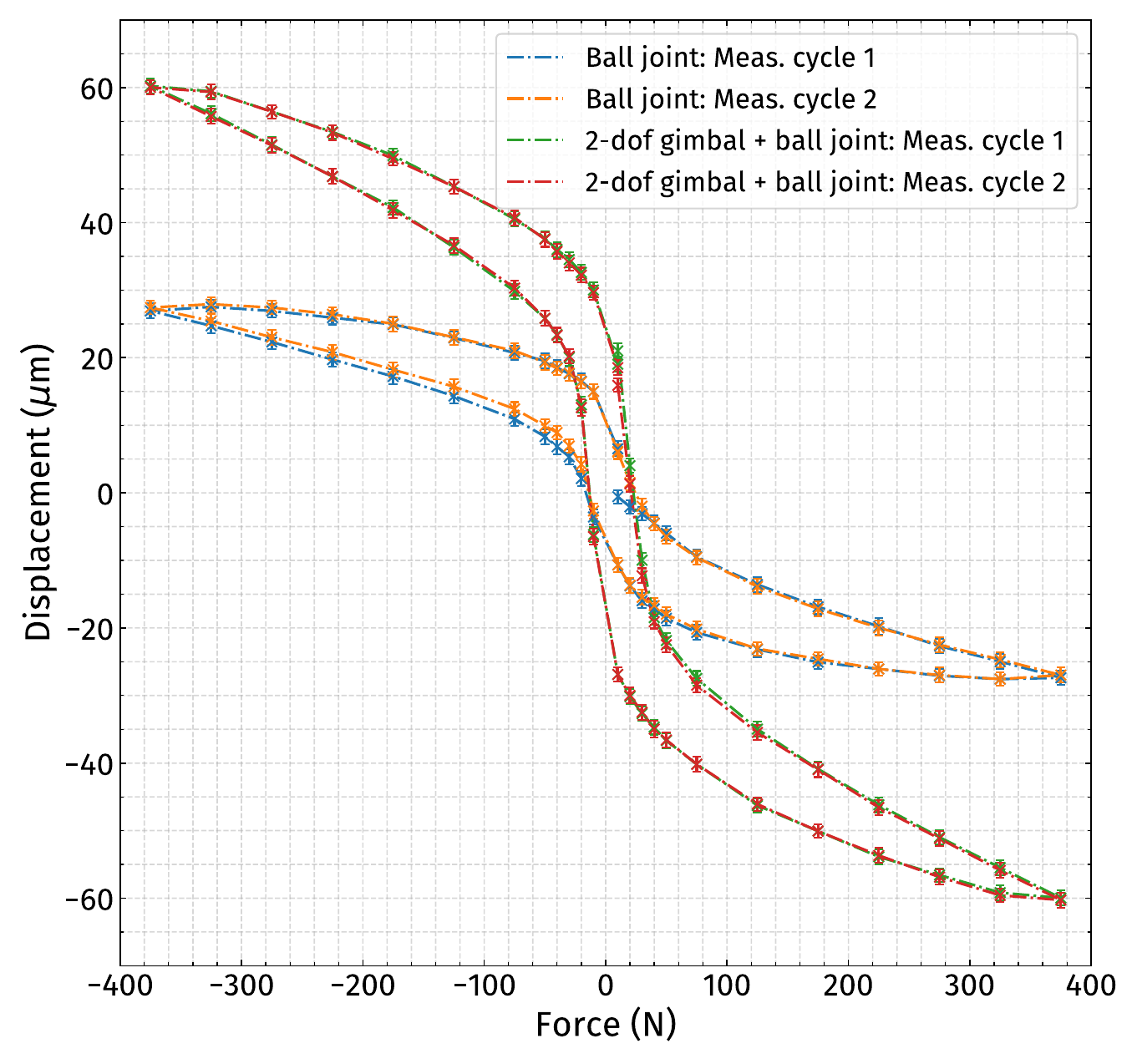}
	\end{subfigure}
	\hfill
	\begin{subfigure}{0.49\textwidth}
		\includegraphics[width=\textwidth]{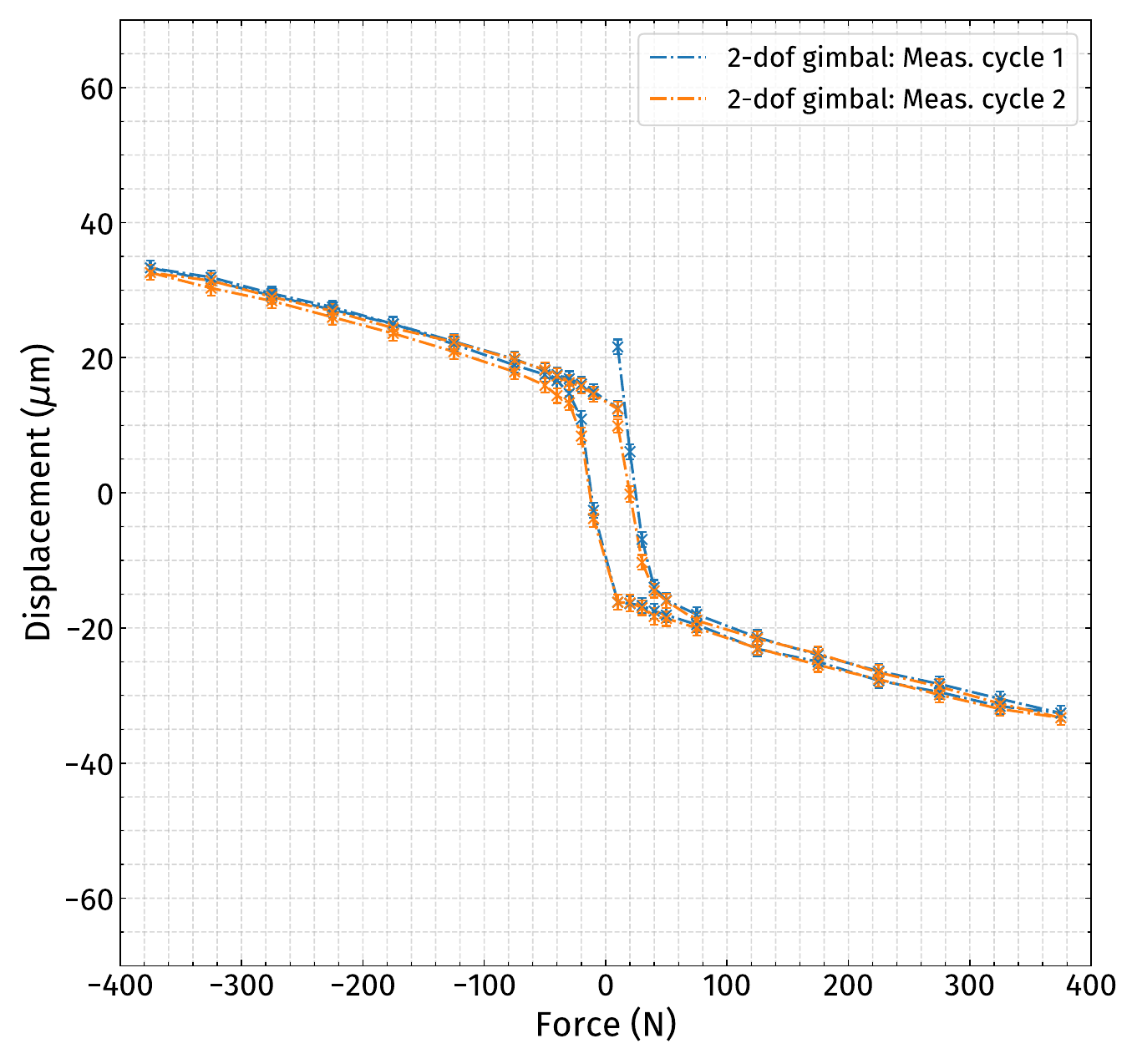}
	\end{subfigure}
	\caption{\textbf{Left}: Force-displacement diagram for a measurement of a dummy system with a 2-dof gimbal and a ball joint (red + green), and with only a ball joint (blue + orange). \textbf{Right}: Force-displacement diagram of the 2-dof gimbal corrected for the effects of the ball joint.}
	\label{sec05prelimres}
\end{figure}

\section{Preliminary Results and Conclusion}
The force-displacement diagram for an exemplary measurement of a dummy system with a ball joint, as well as with a ball joint and 2-dof gimbal is depicted on the left-hand side of \autoref{sec05prelimres}. For this measurement, maximum push ($F_{\text{push}}>0\,$N) and pull forces ($F_{\text{pull}}<0\,$N) of $\pm 375\,$N were chosen. The force steps were selected such as to provide a finer binning around $0\,$N for a better description of the backlash region and larger force steps above and below $\pm 50\,$N to reduce the measurement time. The displacement of a test object is not measured at $0\,$N, because for the measurement of the backlash a finite force $\lessgtr0\,$N is required in order to ensure that all components of the test object exhibiting backlash are in their corresponding mechanical end stops. It should be noted that the chosen measurement force of $\pm 375\,$N is larger than the maximum forces expected during the telescope operations (which are in order of $\pm 130\,$N\footnote{This corresponds to the force on a single actuator expected for the maximum wind speed of $36\,$km/h during observations as specified in the requirements on the MST.}) and was only selected to verify the measurement procedure. \\
The measurement of the dummy system with a ball joint and 2-dof gimbal was corrected by the effects of the ball joint and the resulting force-displacement diagram is shown on the right-hand side of \autoref{sec05prelimres}. The measurement results show clearly that the ball joint introduces a reproducible, force-dependent hysteresis in the force-displacement diagram and ultimately in the positioning accuracy of a complete actuator system. Furthermore, a backlash is introduced by the gimbal due to the movement of the mounting axes in the interface to the mirror support structure. To reduce this latter effect, the installation of set screws is foreseen to improve the fixation of the mounting axes with respect to the interface of the mirror support structure. Additionally, there is hysteresis visible around $0\,$N, which is most likely caused by a not yet considered friction of the sleds of the order of $\sim 20\,$N in the calibration of the force sensor.\footnote{An update of the force sensor calibration taking this friction into account is meanwhile available.} \\
The exemplary measurements indicate that it is challenging but possible to disentangle the backlash introduced by the gimbal from the hysteresis of the ball joint. Additionally, the main conclusion is that the total, force-dependent deformation limits the non-reproducible positioning accuracy, not only the simple force-independent backlash. The qualification of the final MST actuator design is currently pending after improvements to its design.

\end{document}